\documentclass[reprint,
 amsmath,amssymb,
 aps,
]{revtex4-2}

\usepackage{graphicx}% Include figure files
\usepackage{dcolumn}% Align table columns on decimal point
\usepackage{bm}% bold math
\usepackage{xcolor}
\begin{document}

%\preprint{APS/123-QED}

\title{Space-time wave packets violate the universal relationship between angular dispersion and pulse-front tilt}

\author{Layton A. Hall$^{1}$}
\author{Murat Yessenov$^{1}$}
\author{Ayman F. Abouraddy$^{1}$}
%\email{raddy@creol.ucf.edu}
\affiliation{$^{1}$CREOL, The College of Optics \& Photonics, University of Central~Florida, Orlando, FL 32816, USA}

\begin{abstract}
Introducing angular dispersion into a pulsed field tilts the pulse front with respect to the phase front. There exists between the angular dispersion and the pulse-front tilt a universal relationship that is device-independent, and also independent of the pulse shape and bandwidth. We show here that this relationship is violated by propagation-invariant space-time (ST) wave packets, which are  pulsed beams endowed with precise spatio-temporal structure corresponding to a particular form of angular dispersion. We demonstrate theoretically and experimentally that underlying ST wave packets is -- to the best of our knowledge -- the first example in optics of \textit{non-differentiable} angular dispersion, resulting in pulse-front tilt that depends on the pulse bandwidth even at fixed angular dispersion. 
\end{abstract}

\maketitle

Angular dispersion is a condition whereby each wavelength in an optical field travels at a different angle, and is typically induced by dispersive or diffractive devices such as gratings or prisms \cite{Torres10AOP}. Introducing angular dispersion into a pulse results in tilting the pulse front with respect to the phase front \cite{Topp75OC,Schiller80OC}. The relationship between angular dispersion and the tilt of so-called tilted-pulse fronts (TPFs) is universal: the tilt angle is device-independent, and independent of the pulse bandwidth and shape \cite{Hebling96OQE,Fulop10Review}. Moreover, TPFs incur anomalous group-velocity dispersion (GVD) along the propagation direction \cite{Martinez84JOSAA,Fulop10Review}, and are thus useful for dispersion compensation in materials with normal GVD \cite{Szatmari96OL,Torres10AOP}. Angular dispersion and TPFs are versatile tools for traveling-wave excitation \cite{Szatmari90AO,Hebling91JOSAB}, achromatic phase-matching \cite{Martinez89IEEE,Szabo90APB,Dubietis97OL,Schober07JOSAB}, quantum optics \cite{Torres05PRA,Hendrych09PRA}, realizing spatio-temporal solitons \cite{DiTrapani98PRL,Liu00PRE,Wise02OPN}, and generating terahertz radiation \cite{Hebling08JOSAB}.

Here we show that the new class of pulsed beams denoted `space-time' (ST) wave packets \cite{Kondakci16OE,Parker16OE,Kondakci17NP,Yessenov19OPN} that are also undergirded by angular dispersion and are endowed with a symmetrized TPF structure nevertheless violate the universal relationship between angular dispersion and pulse-front tilt. Crucially, previous results regarding TPFs rely fundamentally on the \textit{differentiability} of the angular dispersion with respect to wavelength, and thus the existence of a Taylor expansion for the propagation angle \cite{Hebling96OQE,Fulop10Review,Torres10AOP}. Whereas this assumption is entirely reasonable for fields prepared in a single step via conventional optical components such as gratings and prisms, ST wave packets present -- to the best of our knowledge -- the first example in optics of \textit{non-differentiable} angular dispersion that is readily realized via a two-step procedure \cite{Yessenov19OPN}. In this case a perturbative Taylor expansion is not justified even for narrow bandwidths. As a token of the departure from the typical behavior of TPFs, we find that the pulse-front tilt of ST wave packets depends on the pulse bandwidth, in contrast to conventional TPFs. We elucidate the origin of the non-differentiability of angular dispersion underpinning ST wave packets, derive a new formula relating the pulse-front tilt to the bandwidth and to a wavelength-independent angular descriptor, and confirm these predictions experimentally. By freeing ST wave packets from the constraints associated with conventional angular dispersion, we expect that they can be sculpted to induce arbitrary GVD, which may have important implications for nonlinear and quantum optics. 

\begin{figure*}[t!]
  \begin{center}
  \includegraphics[width=15.6cm]{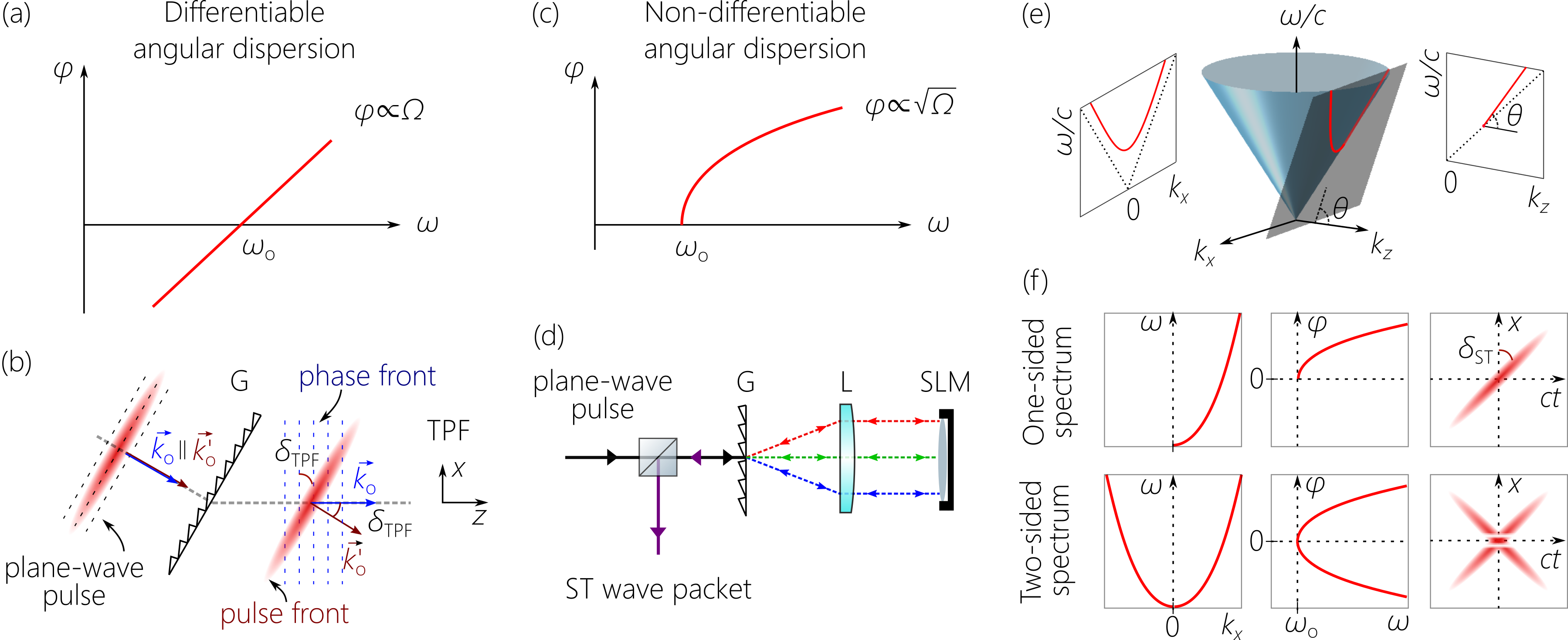}
  \end{center}
  \caption{(a) An example of differentiable angular dispersion $\varphi(\omega_{\mathrm{o}}+\Omega)\!\propto\!\Omega$ as produced in (b) where a plane-wave pulse is incident normally on a grating G. Initially, the phase front and pulse front coincide, and are orthogonal to the parallel vectors $\vec{k}_{\mathrm{o}}$ and $\vec{k}_{\mathrm{o}}'$, respectively. After the grating, $\vec{k}_{\mathrm{o}}'$ is tilted by an angle $\delta_{\mathrm{ST}}$ with respect to $\vec{k}_{\mathrm{o}}$, so the pulse front also makes an angle $\delta_{\mathrm{ST}}$ with respect to the phase front. (c) Non-differentiable angular dispersion for a ST wave packet, $\varphi(\omega_{\mathrm{o}}+\Omega)\!\propto\sqrt{\Omega}$, as produced by the setup in (d); L: lens, SLM: spatial light modulator. (e) Spectral support domain of a ST wave packet in $(k_{x},k_{z},\tfrac{\omega}{c})$ space at the intersection of the light-cone with a plane making an angle $\theta$ with respect to the $k_{z}$-axis. (f) Spectral projection onto the $(k_{x},\tfrac{\omega}{c})$-plane, the angular dispersion $\varphi(\omega_{\mathrm{o}}+\Omega)$, and the spatio-temporal intensity profile of ST wave packets with one-sided and two-sided spatial spectra.}
    \label{Fig:BasicConcept}
\end{figure*}

We start by considering a plane-wave pulse after angular dispersion has been introduced, as manifested by a frequency-dependent propagation angle $\varphi$ typically expanded as follows:
\begin{equation}\label{Eq:AngularExpansion}
\varphi(\omega)=\varphi(\omega_{\mathrm{o}}+\Omega)=\varphi_{\mathrm{o}}+\varphi_{\mathrm{o}}'\Omega+\frac{1}{2}\varphi_{\mathrm{o}}''\Omega^{2}+\cdots;
\end{equation}
here $\varphi_{\mathrm{o}}\!=\!\varphi(\omega_{\mathrm{o}})$, $\varphi_{\mathrm{o}}'\!=\!\tfrac{d\varphi}{d\omega}|_{\omega\!=\!\omega_{\mathrm{o}}}$, $\varphi_{\mathrm{o}}''\!=\!\tfrac{d^{2}\varphi}{d\omega^{2}}|_{\omega\!=\!\omega_{\mathrm{o}}}$, $\omega_{\mathrm{o}}$ is the carrier frequency, and $\Omega\!=\!\omega-\omega_{\mathrm{o}}$ [Fig.~\ref{Fig:BasicConcept}(a)]; here primes denote derivatives with respect to $\omega$, and the subscript `o' quantities evaluated at $\omega\!=\!\omega_{\mathrm{o}}$. This Taylor expansion becomes progressively more accurate as the bandwidth is reduced, and it underpins previous work on spectroscopic devices and TPFs \cite{Hebling94JOSAA,Hebling06JOSAA,Torres10AOP}. The field can be written as follows: $E(\vec{r};t)\!=\!\int\!d\omega\widetilde{E}(\omega)e^{i\xi(\vec{r};\omega)}e^{-i\omega t}$, where $\xi(\vec{r};\omega)\!=\!\vec{k}(\omega)\cdot\vec{r}$, $\widetilde{E}(\omega)$ is the Fourier transform of $E(0;t)$, and the phase $\xi(\vec{r};\omega)=\xi_{\mathrm{o}}(\vec{r})+\xi_{\mathrm{o}}'(\vec{r})\Omega+\tfrac{1}{2}\xi_{\mathrm{o}}''(\vec{r})\Omega^{2}+\cdots$, where $\xi_{\mathrm{o}}(\vec{r})\!=\!\vec{k}_{\mathrm{o}}\cdot\vec{r}$, $\xi_{\mathrm{o}}'\!=\!\vec{k}_{\mathrm{o}}'\cdot\vec{r}$, $\xi_{\mathrm{o}}''\!=\!\vec{k}_{\mathrm{o}}''\cdot\vec{r}$, and $k_{\mathrm{o}}\!=\!\omega_{\mathrm{o}}/c$ \cite{Porras03PRE2}. Aligning the $z$-axis with the direction of  $\omega_{\mathrm{o}}$ (i.e., $\varphi_{\mathrm{o}}\!=\!0$) yields $\xi_{\mathrm{o}}(\vec{r})\!=\!\vec{k}_{\mathrm{o}}\cdot\vec{r}\!=\!k_{\mathrm{o}}z$; that is, the phase front is orthogonal to the $z$-axis. We rewrite the field as the product of a carrier and an envelope $E(\vec{r};t)\!=\!e^{i(k_{\mathrm{o}}z-\omega_{\mathrm{o}}t)}\psi(\vec{r};t)$, where
\begin{equation}
\psi(\vec{r};t)=\int\!d\Omega\widetilde{\psi}(\Omega)e^{-i(t-\xi_{\mathrm{o}}'(\vec{r}))\Omega}e^{i\xi_{\mathrm{o}}''(\vec{r})\Omega^{2}/2};
\end{equation}
here $\widetilde{\psi}(\Omega)$ is the Fourier transform of $\psi(0;t)$. For simplicity, we hold the field uniform along $y$, so the pulse front is orthogonal to $\vec{k}_{\mathrm{o}}'\!=\!k_{\mathrm{o}}\varphi_{\mathrm{o}}'\hat{x}+k_{\mathrm{o}}'\hat{z}$, which makes an angle $\delta_{\mathrm{TPF}}$ with respect to the phase front [Fig.~\ref{Fig:BasicConcept}(b)] and is given in free space by:
\begin{equation}\label{Eq:PulseFrontTiltAngle}
\tan{\delta}_{\mathrm{TPF}}=\omega_{\mathrm{o}}\varphi_{\mathrm{o}}'.
\end{equation}

Crucially, this well-known relationship between angular dispersion $\varphi_{\mathrm{o}}'$ and pulse-front tilt $\tan{\delta_{\mathrm{TPF}}}$ is \textit{universal} \cite{Fulop10Review}: it is independent of the device introducing angular dispersion, and of the pulse bandwidth and shape. The direction of maximum GVD is along the vector $c\vec{k}_{\mathrm{o}}''\!=\!(\omega_{\mathrm{o}}\varphi_{\mathrm{o}}''+2\varphi_{\mathrm{o}}')\hat{x}-\omega_{\mathrm{o}}\varphi_{\mathrm{o}}'^{2}\hat{z}$. The GVD parameter along the propagation axis $z$ is negative-valued, $k_{\mathrm{o},z}''\!=\!-k_{\mathrm{o}}\varphi_{\mathrm{o}}'^{2}$; i.e., angular dispersion always produces anomalous GVD in free space under these considerations \cite{Martinez84JOSAA,Szatmari96OL}.

Implicit in these relationships is the unquestioned assumption that $\varphi(\omega)$ is \textit{differentiable} at $\omega\!=\!\omega_{\mathrm{o}}$, which is necessary to justify the Taylor series in Eq.~\ref{Eq:AngularExpansion}. At first sight, this is a completely reasonable assumption that needs no reconsideration, especially for narrow bandwidths. However, our recent strategy for preparing ST wave packets is capable of incorporating \textit{arbitrarily prescribed} angular dispersion into the field \cite{Kondakci17NP,Kondakci19NC,Yessenov19OE,Yessenov19OPN,Bhaduri19OL,Bhaduri20NP}, which necessitates a reassessment of this assumption.

Consider the angular dispersion in Fig.~\ref{Fig:BasicConcept}(c) where $\varphi(\Omega)\!\propto\!\sqrt{\Omega}$, which is \textit{not} differentiable at $\Omega\!=\!0$, and thus does \textit{not} possess a Taylor expansion such as that in Eq.~\ref{Eq:AngularExpansion}. The functional form of $\varphi(\Omega)$ is free of singularities, is smooth and continuous, and also differentiable away from $\Omega\!=\!0$. Nevertheless, the conventional analysis no longer holds in the vicinity of $\Omega\!=\!0$; e.g., Eq.~\ref{Eq:PulseFrontTiltAngle} is \textit{not} applicable in absence of a well-defined derivative $\varphi_{\mathrm{o}}'$. Although such angular dispersion cannot be produced by a grating or other usual optical components, it can nevertheless be induced by the two-step procedure outlined in Fig.~\ref{Fig:BasicConcept}(d). First, a grating spreads the spectrum in space, but is not relied upon to introduce angular dispersion. Instead, the field is collimated and then a spatial light modulator (SLM) associates to each wavelength a prescribed propagation angle $\varphi(\omega)$. The back-reflected spectrum is reconstituted into a pulse at the grating. In this way, \textit{arbitrary} angular dispersion is realizable. 

Such angular dispersion is necessary for the synthesis of propagation-invariant ST wave packets \cite{Yessenov19OPN} whose spectral support lies at the intersection of the free-space dispersion relationship $k_{x}^{2}+k_{z}^{2}\!=\!(\tfrac{\omega}{c})^{2}$ (the `light-cone') with the spectral plane $\Omega\!=\!(k_{z}-k_{\mathrm{o}})c\tan{\theta}$, where the spectral tilt angle $\theta$ is measured with respect to the $k_{z}$-axis, and is a frequency-\textit{independent} constant angle \cite{Yessenov19PRA}; see Fig.~\ref{Fig:BasicConcept}(e). The resulting ST wave packet propagates rigidly in free space at a group velocity $\widetilde{v}\!=\!c\tan{\theta}$, or an effective group index $\widetilde{n}\!=\!\cot{\theta}$ \cite{Donnelly93ProcRSLA,Kondakci19NC,Yessenov19OE}. In the paraxial regime $(1-\widetilde{n})\tfrac{\Omega}{\omega_{\mathrm{o}}}\!\approx\!\tfrac{k_{x}^{2}}{2k_{\mathrm{o}}^{2}}$, from which we obtain:
\begin{equation}\label{Eq:STWavePacketPropagationDirection}
\sin{\{\varphi(\omega_{\mathrm{o}}+\Omega)\}}=\sqrt{2(1-\widetilde{n})}\frac{\sqrt{\Omega/\omega_{\mathrm{o}}}}{1+\Omega/\omega_{\mathrm{o}}}.
\end{equation}
Taking only positive values for $k_{x}$ and $\varphi$ yields a field with a tilted pulse front whose profile resembles that of a TPF, whereas taking both positive and negative values of $k_{x}$ and $\varphi$ yields a symmetrized TPF structure \cite{Wong17ACSP2,Kondakci19ACSP,Wong20AS}; see Fig.~\ref{Fig:BasicConcept}(f). For small bandwidths $\Delta\Omega\!\ll\!\omega_{\mathrm{o}}$, $\sin{\varphi}\!\approx\!\varphi\!\propto\!\sqrt{\Omega}$, which is \textit{not} a differentiable function at $\Omega\!=\!0$, \textit{independently of the bandwidth} $\Delta\Omega$. We therefore expect that previous results based on the conventional theoretical framework for differentiable angular dispersion do \textit{not} apply to ST wave packets, and for new behaviors to emerge.

\begin{figure}[t!]
  \begin{center}
  \includegraphics[width=8.6cm]{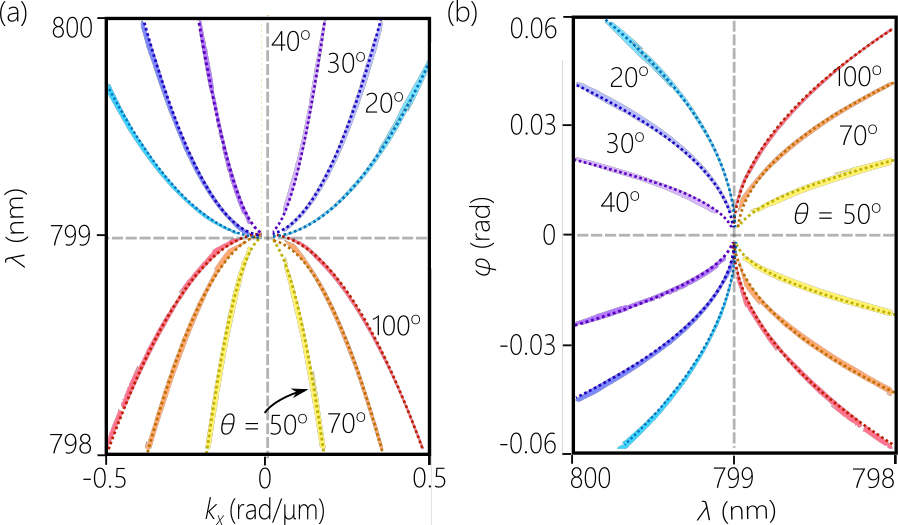}
  \end{center}
  \caption{(a) Measured spatio-temporal spectra of ST wave packets in the $(k_{x},\lambda)$ plane as $\theta$ is tuned, and (b) the corresponding angular dispersion $\varphi(\lambda)$. Colored curves are measurements, and dotted curves are theoretical predictions based on Eq.~\ref{Eq:STWavePacketPropagationDirection}.}
    \label{Fig:MeasuredAngleWavelength}
\end{figure}

We proceed to demonstrate theoretically and experimentally that ST wave packets violate one of the tenets of the universal relationship in Eq.~\ref{Eq:PulseFrontTiltAngle}; namely, the independence of the pulse-front tilt from the pulse bandwidth. The envelope here is $\psi(x,z;t)\!=\!\int\!d\Omega\widetilde{\psi}(\Omega)e^{i\xi(x,z;\Omega)}e^{-i\Omega t}$, where the spectral phase is $\xi(x,z;\Omega)\!=\!k_{\mathrm{o}}\sqrt{\tfrac{\Omega}{\omega_{\mathrm{o}}}}\left\{\eta x+\widetilde{n}\sqrt{\tfrac{\Omega}{\omega_{\mathrm{o}}}}z\right\}$, and $\eta^{2}\!=\!2|1-\cot{\theta}|$. The field amplitude is appreciable when the derivative of this phase changes slowly. The following ansatz minimizes the phase derivative at fixed $z$ and maximizes the pulse amplitude:
\begin{equation}\label{Eq:STPulseFrontTiltAngle}
\tan{\delta}_{\mathrm{ST}}=\pm\sqrt{\frac{|1-\cot{\theta}|}{2\,\Delta\Omega/\omega_{\mathrm{o}}}},
\end{equation}
where the positive sign corresponds to $\theta\!<\!45^{\circ}$, the negative to $\theta\!>\!45^{\circ}$, and $\theta\!=\!45^{\circ}$ corresponds to a plane-wave pulse \cite{Yessenov19PRA} with $\delta_{\mathrm{ST}}\!=\!0$. We therefore predict that the pulse-front tilt for ST wave packets depends on two factors: the wavelength-independent spectral tilt angle $\theta$, and the normalized bandwidth $\Delta\Omega/\omega_{\mathrm{o}}$. To the best of our knowledge, this is the first prediction of its kind based on the non-differentiability of the angular dispersion. Another consequence of the non-differentiable angular dispersion is the GVD-free propagation-invariance of the ST wave packet \cite{Kondakci19NC,Yessenov19OE}, in contrast to the anomalous GVD induced by differentiable angular dispersion \cite{Martinez84JOSAA,Fulop10Review,Torres10AOP}. 

In our experiments, we make use of $\sim\!100$-fs pulses from a mode-locked Ti:sapphire laser (Tsunami, Spectra Physics) at a central wavelength $\approx\!800$~nm directed to the pulse shaper shown schematically in Fig.~\ref{Fig:BasicConcept}(d); see \cite{Kondakci19ACSP,Bhaduri19Optica,Yessenov19OE,Bhaduri20NP} for details. First, a diffraction grating ($1200$~lines/mm) spreads the pulse spectrum in space, and the first diffraction order is collimated with a cylindrical lens of focal length $50$~cm before it impinges on a reflective phase-only SLM (Hamamatsu X10468-02) that imparts to it a two-dimensional phase distribution designed to associate a particular propagation angle $\varphi$ to each wavelength. In this manner, an arbitrary functional form $\varphi(\Omega)$ is realizable, including non-differentiable or even discontinuous mappings, limited only by the spatial and phase resolutions of the SLM and the spectral resolution of the grating. The retro-reflected field from the SLM returns to the grating whereupon the pulse is reconstituted and the ST wave packet formed. The spatio-temporal spectrum, from which we can extract the angular dispersion $\varphi(\Omega)$, is measured by a combination of a grating (to resolve the temporal spectrum) and a lens (to resolve the spatial spectrum). We also acquire the spatio-temporal intensity profile $I(x,z;\tau)\!=\!|E(x,z;\tau)|^{2}$ at a fixed axial plane $z$ by interfering the ST wave packet with a short reference plane-wave pulse while sweeping a relative delay $\tau$; see \cite{Kondakci19ACSP,Bhaduri19Optica,Yessenov19OE,Bhaduri20NP} for details of this experimental procedure.

\begin{figure}[t!]
  \begin{center}
  \includegraphics[width=8.6cm]{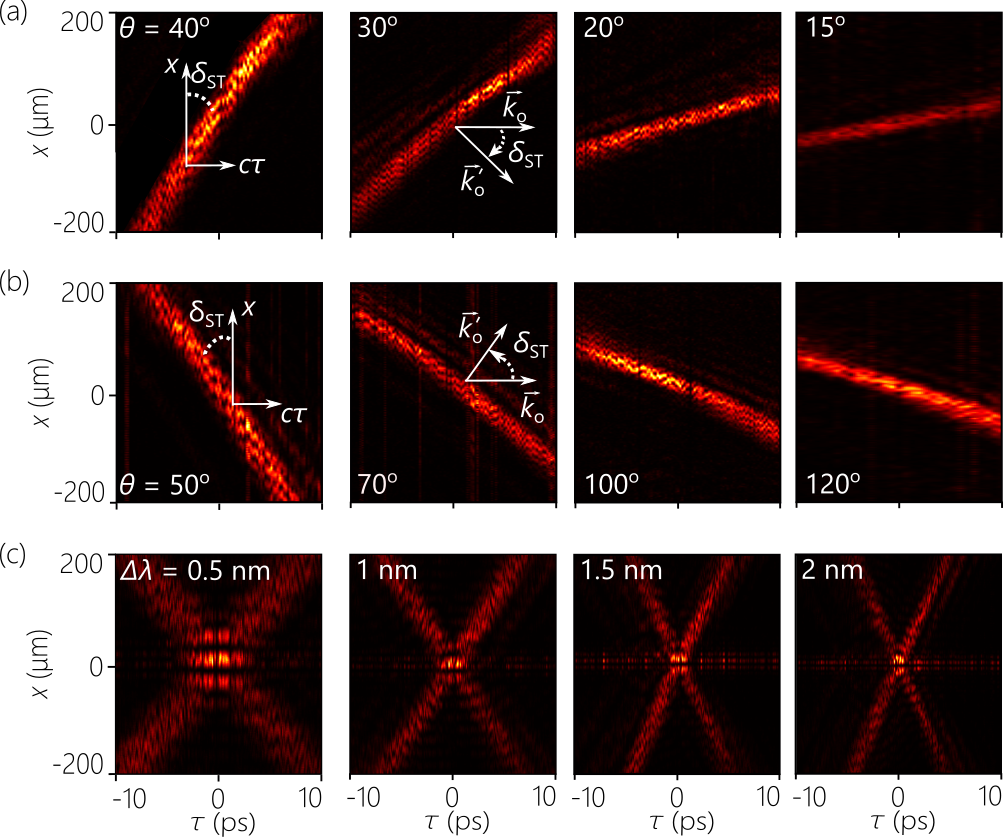}
  \end{center}
  \caption{Measured spatio-temporal intensity profiles $I(x;\tau)$ for ST wave packets at fixed $z$. (a) Profiles while varying $\theta$ at a fixed bandwidth of $\Delta\lambda\!\approx\!1$~nm for single-sided spatial spectra in the subluminal ($\theta\!<\!45^{\circ}$) and (b) superluminal ($\theta\!>\!45^{\circ}$) regimes. (c) Profiles while varying $\Delta\lambda$ at fixed $\theta\!=\!50^{\circ}$.}
    \label{Fig:MeasuredPulseFronts}
\end{figure}

We plot in Fig.~\ref{Fig:MeasuredAngleWavelength}(a) the measured spatio-temporal spectrum in the $(k_{x},\lambda)$-plane while varying the spectral tilt angle $\theta$. Note the opposite curvature of the spectrum when $\theta\!<\!45^{\circ}$ (subluminal, $\widetilde{v}\!<\!c$) and $\theta\!>\!45^{\circ}$ (superluminal, $\widetilde{v}\!>\!c$). From these measurements we extract $\varphi(\lambda)$, which is plotted in Fig.~\ref{Fig:MeasuredAngleWavelength}(b) along with the theoretical expectation from Eq.~\ref{Eq:STWavePacketPropagationDirection}. These measurements fit exactly to a $\sqrt{\Omega}$-dependence, and thus correspond to non-differentiable angular dispersion in the vicinity of $\Omega\!=\!0$.

\begin{figure}[t!]
  \begin{center}
  \includegraphics[width=8.6cm]{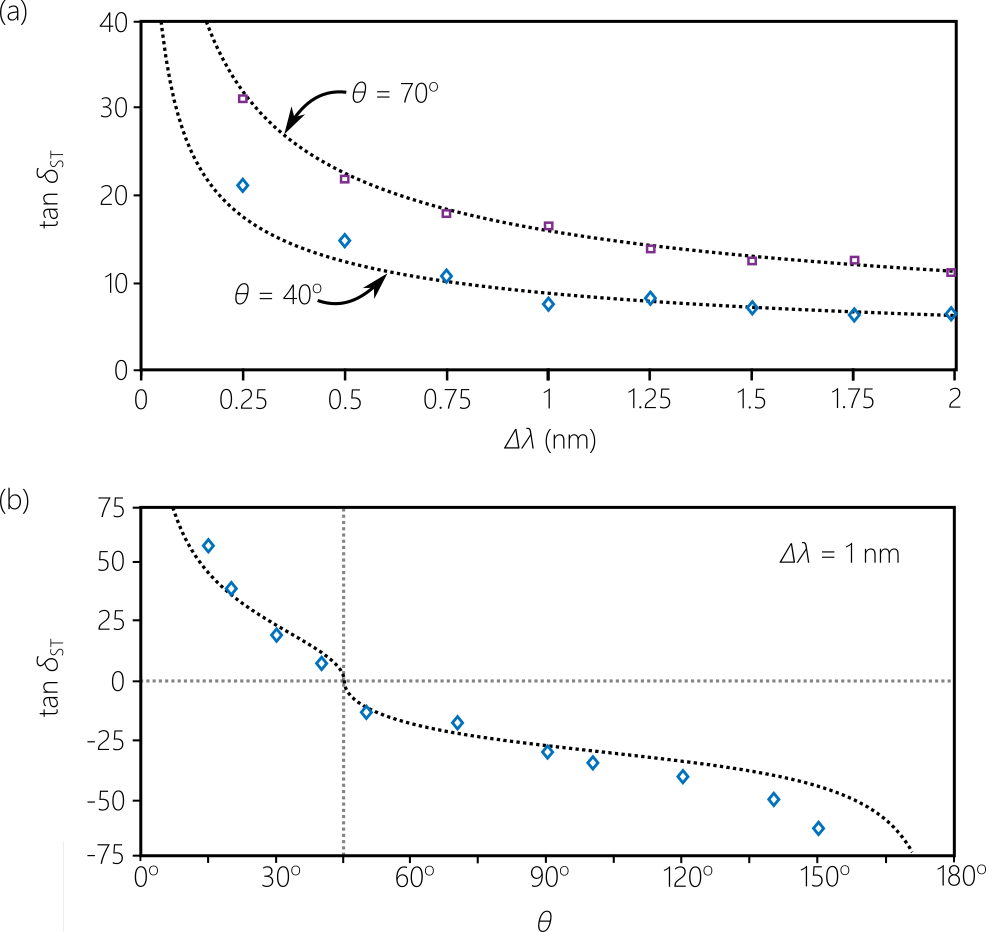}
  \end{center}
  \caption{\small{(a) Measured pulse-front tilt $\tan{\delta_{\mathrm{ST}}}$ for ST wave packets while varying $\Delta\lambda$ at two different values of $\theta$. (b) Measured $\tan{\delta_{\mathrm{ST}}}$ while varying $\theta$ at fixed $\Delta\lambda$. The dashed curves in (a) and (b) are the theoretical expectations $\tan{\delta_{\mathrm{ST}}}\!\propto\!1/\sqrt{\Delta\lambda}$ and $\tan{\delta_{\mathrm{ST}}}\!\propto\!\sqrt{|1-\cot{\theta}|}$, respectively. We remain in the paraxial regime despite the large $\delta_{\mathrm{ST}}$.}}
    \label{Fig:NewRelationship}
\end{figure}

The pulse-front tilt angle $\delta_{\mathrm{ST}}$ is obtained from the measured spatio-temporal intensity profiles $I(x,\tau)$ of the ST wave packets at a fixed $z$; see Fig.~\ref{Fig:MeasuredPulseFronts}. First, we plot in Fig.~\ref{Fig:MeasuredPulseFronts}(a,b) $I(x,\tau)$ for one-sided spatial spectra by suppressing at the SLM the negative values of $k_{x}$ or $\varphi$ shown in Fig.~\ref{Fig:MeasuredAngleWavelength}. These measurements are carried out at a fixed bandwidth of $\Delta\lambda\!\approx\!1$~nm while varying $\theta$. Note the change in the sign of $\delta_{\mathrm{ST}}$ when transitioning from the subluminal regime $\theta\!<\!45^{\circ}$ [Fig.~\ref{Fig:MeasuredPulseFronts}(a)] to the superluminal regime $\theta\!>\!45^{\circ}$ [Fig.~\ref{Fig:MeasuredPulseFronts}(b)]. Next we plot $I(x,\tau)$ for symmetrized two-sided spatial spectra at fixed $\theta\!=\!50^{\circ}$ while varying the bandwidth $\Delta\lambda$ in the range $0.5-2$~nm [Fig.~\ref{Fig:MeasuredPulseFronts}(c)]. When the bandwidth is increased by a factor of 4, the pulse width is reduced by the same factor, whereas $\tan{\delta_{\mathrm{ST}}}$ is reduced by a factor of $\approx\!2$. The observed change in the pulse-front tilt angle $\delta_{\mathrm{ST}}$ with $\Delta\lambda$ is a unique feature of the non-differentiable angular dispersion intrinsic to these ST wave packets. Measurements for $\delta_{\mathrm{ST}}$ at fixed $\theta$ while varying the bandwidth $\Delta\lambda$ are plotted in Fig.~\ref{Fig:NewRelationship}(a). Conversely, measurements of $\delta_{\mathrm{ST}}$ at fixed bandwidth $\Delta\lambda$ while varying $\theta$ are plotted in Fig.~\ref{Fig:NewRelationship}(b). In both case the data fits well the theoretical expectations based on Eq.~\ref{Eq:STPulseFrontTiltAngle}.

The pulsed beams studied here comprise the family of `baseband' ST wave packets because their spatial spectra lie in the vicinity of $k_{x}\!=\!0$ \cite{Yessenov19PRA}; i.e., $\varphi(\omega_{\mathrm{o}})\!=\!0$. The well-known family of X-waves \cite{Lu92IEEEa,Saari97PRL,FigueroaBook14}, whose members are also propagation-invariant wave packets in free space, is \textit{free} of angular dispersion because $k_{x}\!\propto\!\omega$ and $\varphi(\omega)\!=\!\varphi_{\mathrm{o}}\!\neq\!0$ \cite{Turunen10PO}. The lack of angular dispersion implies that X-waves are GVD-free and that they are amenable to the conventional perturbative formulation \cite{Porras03PRE2}. Because the well-known result in Ref.~\cite{Martinez84JOSAA} -- that angular dispersion necessarily yields anomalous GVD -- is based on a perturbative expansion of $\varphi(\Omega)$, it does not hold here. We therefore expect that the non-differentiable angular dispersion underpinning baseband ST wave packets can allow for synthesizing new wave packets that encounter arbitrarily specified GVD in free space. This can be achieved by tuning the spectral dependence of the axial wave number $k_{z}$ away from the linear relationship $\Omega\!=\!(k_{z}-k_{\mathrm{o}})\widetilde{v}$. We are currently carrying out experiments along these lines.

In conclusion, we have shown that underpinning ST wave packets is the first example of non-differentiable angular dispersion in optics. The well-established relationship between angular dispersion and pulse-front tilt is founded upon the differentiability of the propagation angle. This result is not valid for ST wave packets whose underlying angular dispersion is non-differentiable. We have experimentally verified a new formula for ST wave wave packets whose pulse-front tilt depends on a wavelength-independent parameter (the spectral tilt angle $\theta$), and -- surprisingly -- the pulse bandwidth, in contradistinction to conventional TPFs whose tilt angle is bandwidth-independent. These results elucidate a fundamental difference between ST wave packets and TPFs despite both being consequences of angular dispersion in pulsed fields, and thus potentially paves the way to new avenues for implementing arbitrary GVD compensation schemes in waveguides \cite{Shiri20NC}, surface plasmon polaritons \cite{Schepler20ACSP}, and nonlinear optics. 

\section*{Funding}
U.S. Office of Naval Research (ONR) contract N00014-17-1-2458 and ONR MURI contract N00014-20-1-2789.

\section*{Disclosures}
The authors declare no conflicts of interest.

\bibliography{diffraction}

\end{document}